\documentclass{aa}
\usepackage{graphicx} 
\usepackage{txfonts}
\usepackage{natbib}
\bibpunct{(}{)}{;}{a}{}{,}

\begin{document}
  \title {V2051 Oph after superoutburst: \\out-of-plane material and the superhump light source\thanks{Based on data collected at the ESO La Silla Observatory using the NTT (ESO proposal 63.H-0010(A)).}}

  \author {Christina Papadaki\inst{1,2}, Henri M.J. Boffin\inst{3}, Danny Steeghs\inst{4} \& Linda Schmidtobreick\inst{1}}
  \institute{ESO, Casilla 19001, Santiago19, Chile
    \and Vrije Universiteit Brussel, Pleinlaan 2, 1050 Brussels, Belgium 
    \and ESO, Karl-Schwarzschild-Str. 2,
    85738 Garching, Germany
     \and Department of Physics, University of Warwick, UK}
  \date{Received date / Accepted date}
  
  \abstract  
      {}
      {We performed a detailed spectroscopic analysis of the dwarf nova \object{V2051 Oph} at the end of its 1999 superoutburst. We studied and interpreted the simultaneous behaviour of various emission lines.}
      {We obtained high-resolution echelle spectroscopic data at ESO's NTT with EMMI, covering the spectral range of 4000--7500$\AA$. The analysis was performed using standard IRAF tools. The indirect imaging technique of Doppler tomography was applied, in order to map the accretion disc and distinguish between the different emission sources. }
      {The spectra are characterised by strong Balmer emission, together with lines of \ion{He}{i} and the iron triplet \ion{Fe}{ii} {42}. All lines are double-peaked, but the blue-to-red peak strength and central absorption depth vary. The primary's velocity was found to be 84.9\,$\rm km \, s^{-1}$. The spectrograms of the emission lines reveal the prograde rotation of a disc-like emitting region and, for the Balmer and \ion{He}{i} lines, an enhancement of the red-wing during eclipse indicates a bright spot origin. The modulation of the double-peak separation shows a highly asymmetric disc with non-uniform emissivity. This is confirmed by the Doppler maps, which apart from the disc and bright spot emission also indicate an additional region of enhanced emission in the 4$^{th}$ quadrant (+$\rm V_x$,$-\rm V_y$), which we associate with the superhump light source. Given the behaviour of the iron triplet and its distinct differences from the rest of the lines, we attribute its existence to an extended gas region above the disc. Its origin can be explained through the fluorescence mechanism.}
      {}
      
      \titlerunning{V2051 Oph after superoutburst}
      \maketitle
 %%%%%%%%%%%%%%%%%%%%%%%%%%%%%%%%%%%%%%%%%%%%%%%%%%%%%%%%%%%%%%%%%%%%%%%%%%%%%
 %%%%%%%%%%%%%%%%%%%%%%%%%%%%%%%%%%%%%%%%%%%%%%%%%%%%%%%%%%%%%%%%%%%%%%%%%%%%%
%%%%%%%%%%%%%%%%%%%%%%%%%%%%%%%%%%%%%%%%%%%%%%%%%%%%%%%%%%%%%%%%%%%%%%%%%%%%%
      \begin{figure*}
	\centering
	\includegraphics[width=18.5cm]{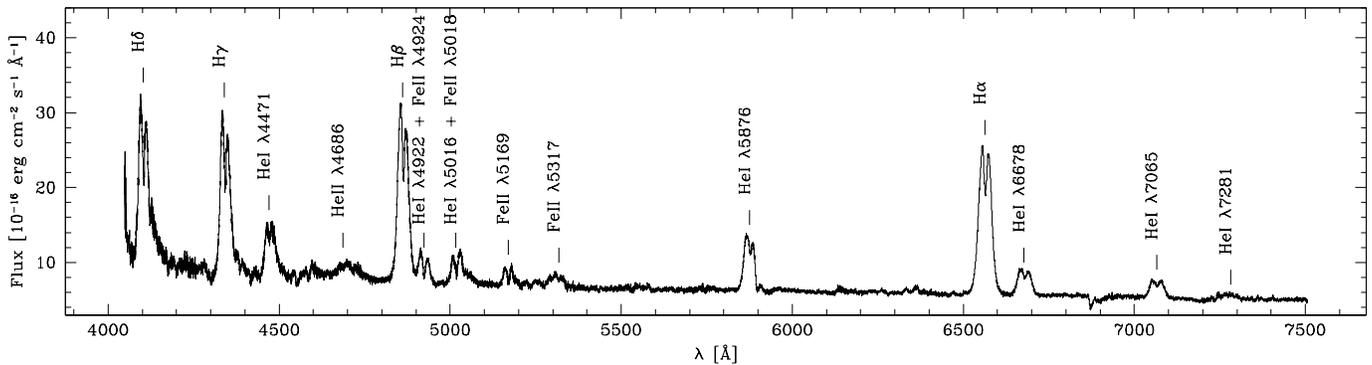}
	\caption{\object{V2051 Oph} average out-of-eclipse spectrum. All lines are double-peaked with prominent emission from the Balmer and \ion{He}{i} lines. \ion{He}{ii} and \ion{Fe}{ii} lines are also in emission, albeit weaker.}
	\label{f_avgsp_v2051oph}
      \end{figure*}
%%%%%%%%%%%%%%%%%%%%%%%%%%%%%%%%%%%%%%%%%%%%%%%%%%%%%%%%%%%%%%%%%%%%%%%%%%%%%

      \section{Introduction}
      The SU UMa stars are cataclysmic variables (CVs) belonging to the sub-category of dwarf novae (DN). Like all CVs they consist of a red dwarf (secondary) that fills its Roche lobe and orbits around a white dwarf (primary). The primary accretes matter from the secondary through an accretion disc (AD). The collision between the accretion stream and the rim of the AD results in the dissipation of energy and the formation of a ``bright spot'' (BS). The DN are characterised by two brightness states, quiescence and outburst of typically 2--5\,mag. The main characteristics of the SU UMa stars are that they undergo superoutbursts, where the system's brightness increases even more than in a normal outburst, and have orbital periods below or, in some cases, in the CV period gap of 2--3\,h. For an extensive review of DN and SU UMa stars, in particular, see \citet{war95}, \citet{osa01} and references therein.

      \object{V2051 Oph} (RA 17:08:19.09, DEC -25:48:30.8, J2000) was discovered by \citet{san72} and has ever since been an object of great interest and continuous study. Although a lot of questions have been raised on its classification as a low-field polar \citep{war87} or a DN, it is now widely believed to belong to the SU UMa subtype of DN. It has a short period of 1.5\,h \citep{war83} and shows recurrent but sparse outbursts \citep{war83,war87} and superoutbursts \citep{kat01}. Its visual magnitude is $\approx$15\,mag in quiescence \citep{war83}, 13.2--14.6\,mag in outburst and 11.6--12.9 in superoutburst \citep{kat01}. A positive superhump, i.e. a modulation originating from the beating between the prograde precession of the AD and the orbital motion of the binary, has also been detected by \citet{kiy98} and confirmed by \citet{pat03}. The light curves of the latter were obtained by G. Garradd\footnote{The light curves can be retrieved at ftp://ftp.kusastro.kyoto-u.ac.jp/pub/vsnet/DNe/V2051\_Oph/1999lc10.gif.} and show a clear hump before each eclipse, indicating the presence of a strong BS.  The system's high inclination $i\approx83\degr$ and therefore deep eclipses, along with follow-up observations during an unusually low-brightness state ($B\approx 16.2\,\rm mag$), allowed \citet{bap98} to accurately define the binary parameters.

      Compared to other SU UMa systems, \object{V2051 Oph} displays a variety of eclipse morphologies and shows humps at various phases and irregular occurrences \citep{war83,war87,vri02}. This behaviour was the cornerstone of the polar scenario, especially since no superoutbursts had been observed up to that point. Another characteristic of the system is the large amplitude flickering activity during quiescence. It reaches 30\% of the total light in the optical and occurs at all phases apart from eclipse \citep{coo83}. The flickering source was studied through eclipse mapping by \citet{bap04}, who found two different dominant contributions: (1) low-frequency flickering, mainly arising from an overflowing gas-stream and (2) high-frequency flickering, produced in the AD.  

      In the first spectroscopy studies, \object{V2051 Oph} clearly showed Balmer, \ion{He}{i} and \ion{He}{ii} lines in emission \citep{wil83}. Profile distortion during eclipse, characteristic of a prograde rotation of a disc-like emission line region \citep{coo83}, was also evident. A study during quiescence, covering 4100--5000\,\AA\, \citep{wat86}, showed the blue peak to be stronger than the red, in all emission lines, and the peak separation in H$\gamma$ to be larger than that in H$\beta$, pointing to a more central location of the former. UV spectra obtained from the same study were similar to that of DN in quiescence. \cite{war87} obtained the first spectra during decline from an outburst (1\,mag higher than quiescence), and found that the separation and widths of the lines to be similar to the ones published by \cite{wat86} during quiescence. However, there was a distinct difference between the set of spectra during the two states: The V/R variations (flux ratio of the blue to the red wings) around the orbit were in almost exact anti-phase, indicating a large change in the distribution of the line-emitting regions \citep{war87}. It has to be noted that in both cases no evidence for BS emission nor the resulting S-wave was found. The first Doppler maps of the H$\beta$ and \ion{He}{i} $\lambda$4471\,\AA\, lines, according to the back-projection technique, are presented in \citet{kai94}. The trailed spectra of the lines confirmed the double-peaked emission and rotational disturbance, while the maps showed the ring-shaped emission pattern of the disc. \citet{ste01} were the first to detect and resolve dwarf nova oscillations in the emission lines. Coherent oscillations were also detected in the continuum. The same study showed that the disc is highly asymmetric with a stronger contribution from its blue-shifted side. Recently, \citet{sai06} reported a study of the spectra and the AD while the system was in an unusual low brightness state during 1996. Among others, they proposed a model of a vertically extended disc with the line emission  arising from the upper atmospheric layers, gas stream overflow, an extended gas region above the disc and a non-negligible opening angle of the disc.

The aim of this work is to accomplish a detailed spectroscopic study of \object{V2051 Oph}. 
Albeit as mentioned above, \object{V2051 Oph} shows some rather unique behaviour, it is also a very useful representative of the class of dwarf novae, and as such, serves to understand the physics at play in these systems better. In particular, one can investigate the possible presence of stream-disc overflow or of matter outside of the disc's plane. Moreover, our unique set of spectra, with its broad wavelength coverage and high resolution gives us the opportunity to simultaneously study the behaviour of the various emission lines and, through the indirect imaging technique of Doppler tomography, to map for the first time the emission distribution of the AD in detail. This study thus also serves to show the merit of echelle spectroscopy-based Doppler tomography.
%This work is part of a project initiated by HMJB and D. Steeghs.  

This paper is organised as follows: in Sect. 2 we present our observations and data reduction technique, in Sect. 3 we perform the spectral analysis and Sect. 4 deals with the application and results of Doppler tomography. Last, Sect. 5 is devoted to the discussion of our two principal results, while Sect. 6 gives a summary.    

      \section{Observations and data reduction}
      We observed \object{V2051 Oph} on the $4^{th}$ of August 1999, at ESO's NTT with the EMMI spectrograph in its echelle, REMD mode \citep{dek86}. The echelle grating \#9 and cross-disperser \#3 resulted in 21 orders covering the wavelength range 4000--7500\,\AA \,with a dispersion of 0.09 to 0.17\,\AA/pix from blue to red, respectively. In total, the system was observed for 4\,h, 2.7 orbits were covered and 23 spectra, each with an exposure time of 300\,s, were obtained. 

      The spectra were reduced following the standard reduction procedures in {\it IRAF}\footnote{{\it IRAF} is distributed by the National Optical Astronomy Observatory, which is operated by the Association of Universities for Research in Astronomy, Inc., under cooperative agreement with the NSF.}. The order extraction, wavelength calibration, flux calibration and order merging were performed by the {\it echelle} package and the reduction task {{\it doecslit} in {\it IRAF}. The spectrophotometric standard \object{LTT 7379} \citep{ham92,ham94} was observed during the same night and served for the flux calibration. The spectra were also corrected for cosmic rays.

      \object{V2051 Oph} was observed 2 days after the end of the 1999 superoutburst \citep{kat01}, when its brightness rapidly fell to approximately that of quiescence. Nevertheless, photometry conducted by \citet{pat03} during the same night, revealed clear superhumps, showing therefore that the disc was still large enough to precess and had not fully recovered from outburst.

\begin{table}
  \caption {Characteristics of emission lines.}
  \label{t_lin_v2051oph}
  \centering
  \begin{tabular}{llll}
    \hline\hline
    & FWZI & EW & Integrated Flux\\
    & [km\,s$^{-1}$] & [\AA] & [$10^{-13}$ erg cm$^{-2}$ s$^{-1}$]\\
    \hline
    H$\delta$ & 5695 & 60.3 & 6.5\\
    H$\gamma$ & 5945 & 78.4 & 6.7\\
    \ion{He}{i}  $\lambda$4471 & 4630 & 27.3 & 2.3\\
    H$\beta$  & 5395 & 94.8 & 7.5\\
    \ion{He}{i}  $\lambda$4922+\ion{Fe}{ii} $\lambda$4924 & 3199 & 11.4 & 0.88\\
    \ion{He}{i} $\lambda$5016+\ion{Fe}{ii} $\lambda$5018 & 3296 & 14.5 & 1.1\\
    \ion{Fe}{ii} $\lambda$5169 & 3472 & 9.8 & 0.68\\
    \ion{He}{i}  $\lambda$5876 & 3043 & 35.9 & 2.3\\
    H$\alpha$ & 4428 & 140.5 & 8.1\\
    \ion{He}{i}  $\lambda$6678 & 3503 & 26.3 & 1.5\\
    \ion{He}{i}  $\lambda$7065 & 3223 & 19.0 & 1.0\\
    \hline
  \end{tabular}
\end {table}

\begin{table}
  \caption {V2051 Oph emission line wing velocity parameters.}
  \label{t_diag_v2051oph}
  \centering
  \begin{tabular}{lllll}
    \hline\hline
    & $K_1$ & $\gamma$ & $\phi_{0}$ & d\\
    & [km\,s$^{-1}$] & [km\,s$^{-1}$] & & [\AA] \\
    \hline
    H$\delta$ & 53(17) & 3(12) & 0.22(6)&35\\
    H$\gamma$ & 95(21) & 61(9)& 0.20(2)&35\\
    \ion{He}{i}  $\lambda$4471 & 134(26) & 93(18) & 0.16(3)&39\\
    H$\beta$  & 99(14) & -7(9) & 0.19(3)&42\\
    \ion{He}{i}  $\lambda$4922 + \ion{Fe}{ii} $\lambda$4924 & 113(54) & 0(38) & -0.14(10)&19 \\
    \ion{He}{i} $\lambda$5016 + \ion{Fe}{ii} $\lambda$5018 & 190(41) & 166(30)& 0.15(4)&34 \\
    \ion{Fe}{ii} $\lambda$5169 & 58(26) & -2(18) & 0.01(10)& 42\\
    H$\alpha$ & 99(19) & -13(9) & 0.17(2)&55\\
    \ion{He}{i}  $\lambda$6678 & 62(14) & -14(10) & 0.11(5)&53 \\
    \ion{He}{i}  $\lambda$7065 & 84(18)& 2(12) & 0.23(5)&52 \\
    \hline
  \end{tabular}
\end{table}

\section{Spectral analysis}

%%%%%%%%%%%%%%%%%%%%%%%%%%%%%%%%%%%%%%%%%%%%%%%%%%%%%%%%%%%%%%%%%%%%%%%%%%%%%
      \begin{figure*}
	\centering
	\includegraphics[width=18cm]{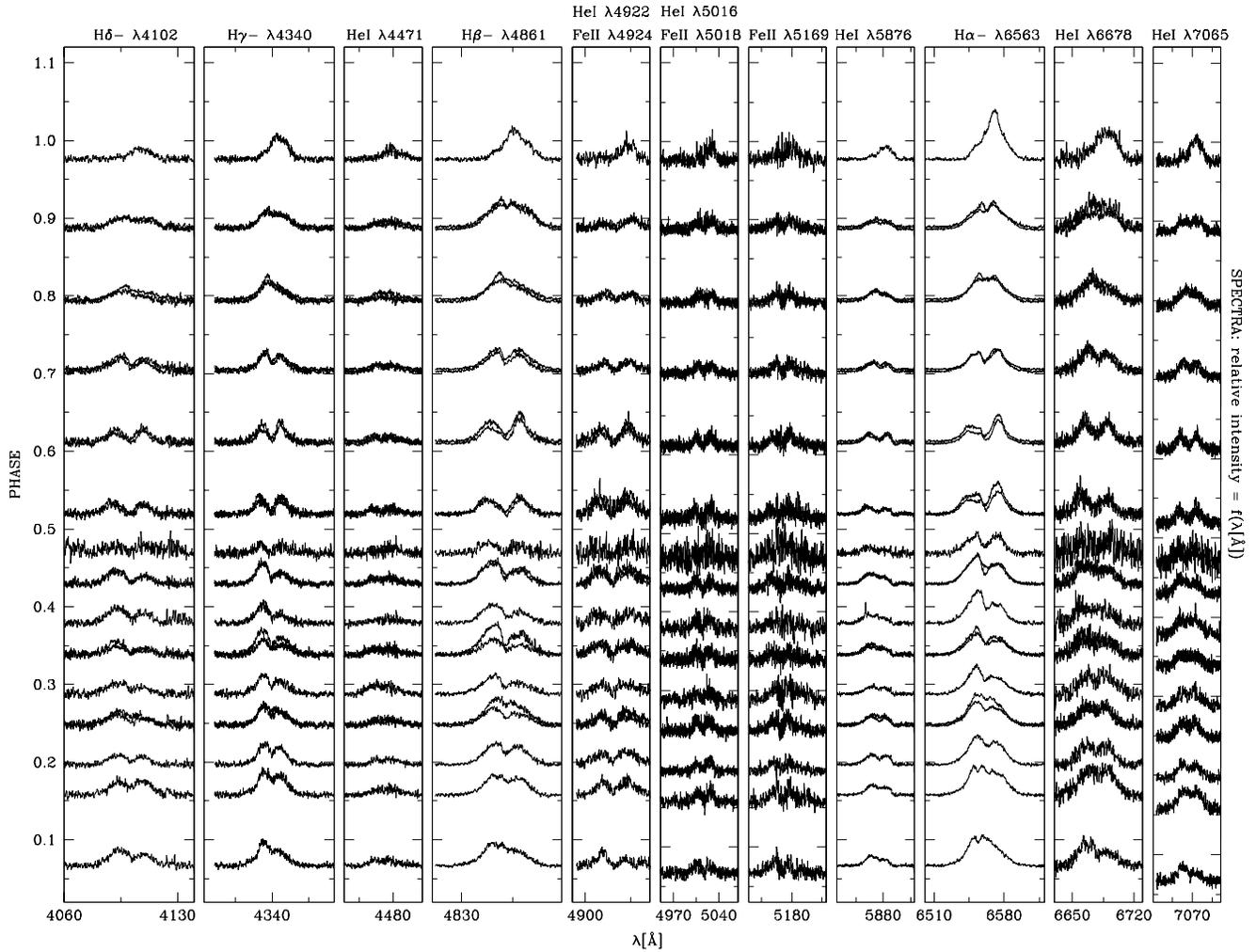}
	\caption{\object{V2051 Oph} trailed spectra of the 11 most prominent lines. The spectra have been ordered according to phase.}
	\label{f_trsp_v2051oph}
      \end{figure*}
%%%%%%%%%%%%%%%%%%%%%%%%%%%%%%%%%%%%%%%%%%%%%%%%%%%%%%%%%%%%%%%%%%%%%%%%%%%%%

%% %%%%%%%%%%%%%%%%%%%%%%%%%%%%%%%%%%%%%%%%%%%%%%%%%%%%%%%%%%%%%%%%%%%%%%%%%%%%%
%%       \begin{figure}
%% 	\centering
%% 	\includegraphics[width=9cm]{v2051oph_lcs.eps}
%% 	\caption{\object{V2051 Oph} continuum and emission-line light curves. The flux and equivalent width variation is shown in each panel's upper and lower graphs, respectively.}
%% 	\label{f_eqw_v2051oph}
%%       \end{figure}
%% %%%%%%%%%%%%%%%%%%%%%%%%%%%%%%%%%%%%%%%%%%%%%%%%%%%%%%%%%%%%%%%%%%%%%%%%%%%%%     

%%%%%%%%%%%%%%%%%%%%%%%%%%%%%%%%%%%%%%%%%%%%%%%%%%%%%%%%%%%%%%%%%%%%%%%%%%%%%
      \begin{figure*}
	\centering
	\includegraphics [width=15cm]{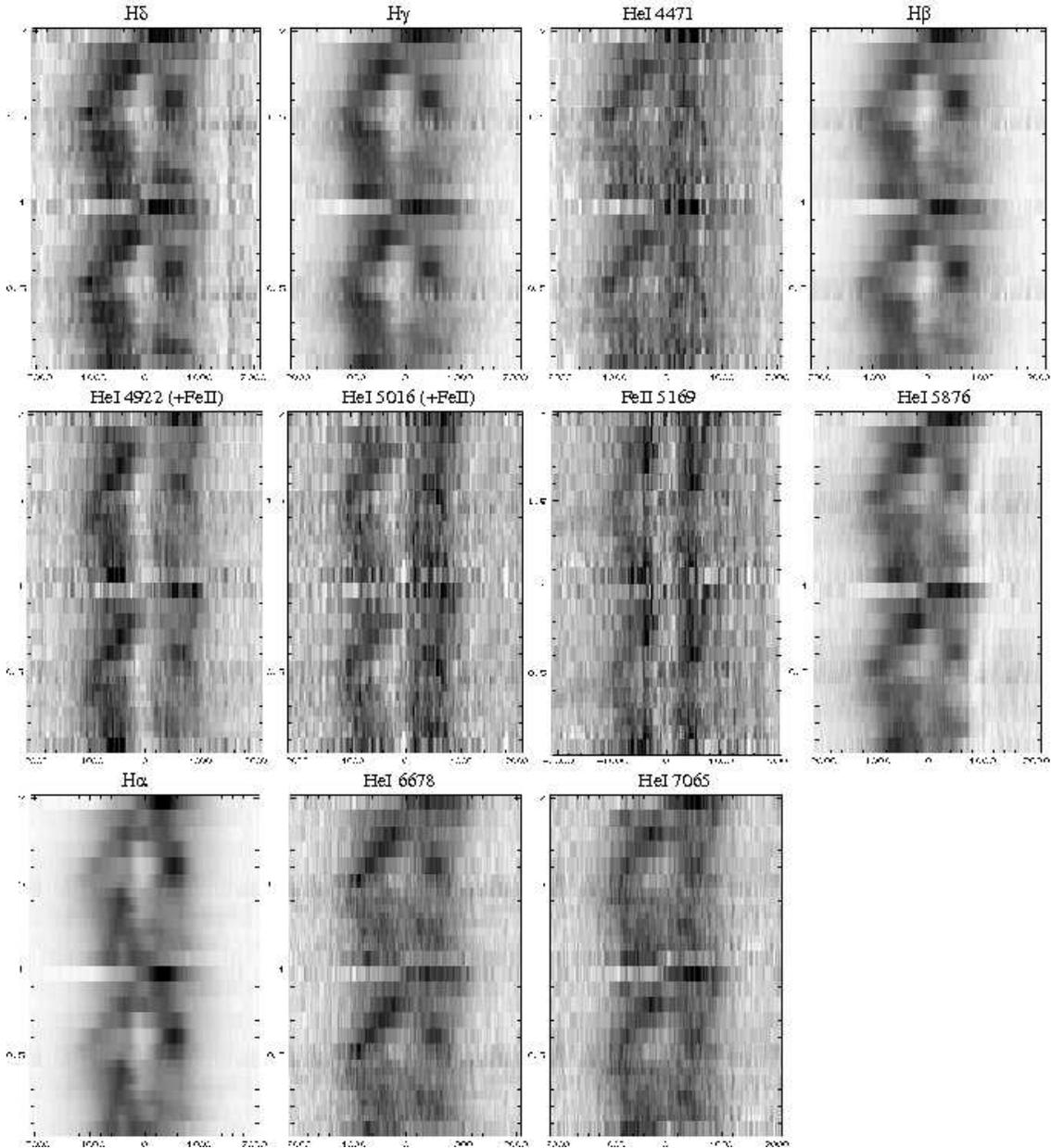}
	\caption{\object{V2051 Oph} trailed spectrograms. The x-axis represents the velocity in km\,s$^{-1}$ and the y-axis the orbital phase. The grayscale contrast has been adjusted to emphasise between emission (black ) and absorption (white).}
	\label{f_spectro_v2051oph}
      \end{figure*}
%%%%%%%%%%%%%%%%%%%%%%%%%%%%%%%%%%%%%%%%%%%%%%%%%%%%%%%%%%%%%%%%%%%%%%%%%%%%% 

The average out-of-eclipse spectrum is shown in Fig.~\ref{f_avgsp_v2051oph}. All lines are in emission and appear double-peaked. The prominent ones are marked on the same figure. The spectrum is characterised by strong Balmer emission on a slightly blue continuum, together with lines of \ion{He}{i} and a weak line of \ion{He}{ii} $\lambda4686$. Lines of the heavier element, \ion{Fe}{ii}, are also present. All Balmer and \ion{He}{i} lines are asymmetric and show a stronger blue peak, although in the Balmer lines the difference in strength is more pronounced. Only in the case of \ion{He}{i} $\lambda$5016 is the red peak stronger but that line is blended with \ion{Fe}{ii}. As for the unblended \ion{Fe}{ii} (5169\,\AA) line, it shows a stronger red-peak. The Balmer decrement from H$\alpha$ to H$\delta$ is 1:0.67:0.56:0.43, showing a mix of optically thick and thin emission. We also searched for orbital variations of the Balmer decrement, but none were evident. Table~\ref{t_lin_v2051oph} shows the full width at zero intensity (FWZI), equivalent width (EW) and integrated flux of the 11 most prominent lines on which we will focus. 

The three  \ion{Fe}{ii} lines ($\lambda$4924, $\lambda$5018 and $\lambda$5169) compose the iron triplet \ion{Fe}{ii} {42}. \ion{Fe}{ii} $\lambda$4924 and $\lambda$5018 are blended with \ion{He}{i} lines, which are fairly strong in our spectra. However, the existence of the \ion{Fe}{ii} $\lambda$5169, the deeper central absorption of all \ion{Fe}{ii} lines (compared to the rest) and the inversion of blue-to-red peak strength in all \ion{Fe}{ii} lines  (except the one at $\lambda$4924) make us conclude that \ion{Fe}{ii} has a non-negligible contribution in the blended lines. The only case where \ion{Fe}{ii} could be less dominant is in the blend with \ion{He}{i} $\lambda$4922. A fourth \ion{Fe}{ii} emission line is seen at $\lambda$5317 but is too weak to be included in further analysis.

A red-blue wing asymmetry is observed in the lines on the blue part of the spectrum, in Fig.~\ref{f_avgsp_v2051oph}. To be sure that spectra belonging to incomplete orbits are not responsible for the asymmetries, the average spectra of the individual full orbits were calculated as well, but the pattern persisted. We also checked if the merging of the echelle orders could be responsible, but no significant order mismatches were evident. However, the order merging is responsible for the higher noise of the red wing of H$\delta$ (Fig.~\ref{f_avgsp_v2051oph}, Fig.~\ref{f_trsp_v2051oph}). The asymmetry could therefore originate either from blended lines or from an enhancement of the continuum. The first scenario could possibly be the cause of the asymmetry for H$\delta$ and H$\gamma$, owing to the proximity of the \ion{He}{i} lines at 4121 and 4388\,\AA, respectively. However, the blend of \ion{He}{i} $\lambda$5016 with \ion{Fe}{ii} $\lambda$5018, which shows the same red-wing asymmetry, has no emission line in the proximity interfering with the red wing. The same applies to \ion{He}{i} $\lambda$4471. For longer wavelengths the lines appear symmetric. The same behaviour was indicated by \citet{wat86} and \citet{ste01}, giving further credit to its existence.

The trailed spectra, calculated according to the following orbital ephemeris \citep{bap03},  

%%%%%%%%%%%%%%%%%%%%%%%%%%%%%%%%%%%%%%%%%%%%%%%%%%%%%%%%%%%%%%%%%%%%%%%%%%%%%
      \begin{figure}
	\centering
	\includegraphics[width=9cm]{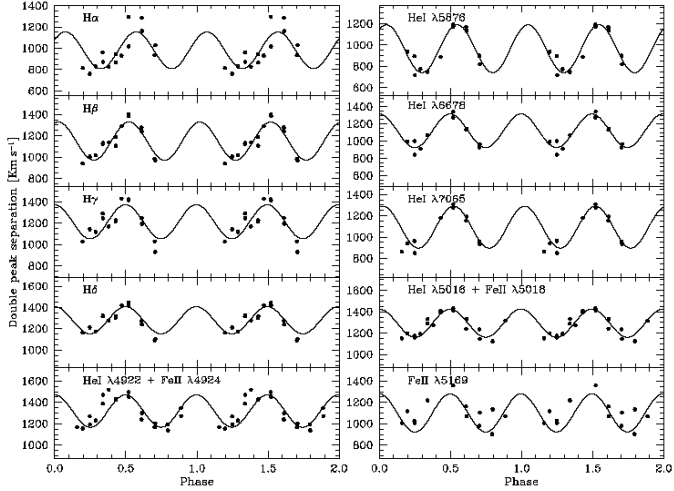}
	\caption{Modulation of the double-peaks separation. Superimposed is the dominant periodicity of $\rm P_{\rm orb}/2$.}
	\label{f_dp_v2051oph}
      \end{figure}
%%%%%%%%%%%%%%%%%%%%%%%%%%%%%%%%%%%%%%%%%%%%%%%%%%%%%%%%%%%%%%%%%%%%%%%%%%%%%     

\begin {equation}
  T_{0}[HJD]=2443245.976900+0\fd062427860E
\end {equation}
where $T_{0}$ is the time of mid-eclipse, are folded on the orbital period $\rm P_{\rm orb}$ and shown in Fig.~\ref{f_trsp_v2051oph}. Some of the spectra overlap since consecutive cycles have been covered. For H$\alpha$, H$\beta$ and H$\gamma$ the intensity between two same phases belonging to consecutive cycles changes somewhat, possibly indicating a modulation on a different period, e.g. a superhump.  
We have however verified that this does not affect the trailed spectrograms nor the Doppler maps, and in the following analysis we will thus use the full set of data.

The diagnostic diagrams, encompassing the double Gaussian fitting technique described by \citet{sha83}, were used to estimate the projected primary velocity $K_1$ and the systemic velocity $\gamma$. When using the information from all emission lines, $K_1$ was found to be 98.73$\pm$25.10 km\,s$^{-1}$. However, if we avoid the weaker lines which give higher discrepancies in the diagnostic diagrams, we obtain a refined value of 78.51$\pm$18.58 km\,s$^{-1}$. This value is in good agreement with the one of 83$\pm$12 km\,s$^{-1}$ from \citet{bap98} and 91 km\,s$^{-1}$ from \citet{wat86}. The first one was predicted by a photometric model based on measurements of contact phases in the eclipse light curve, while the second one was derived from the radial velocity curves of H$\beta$ and H$\gamma$. As for $\gamma$, even though it varied around 0 for most of the lines, it had greater values for some of them. Discrepancies of $\gamma$ values between the emission lines of \object{V2051 Oph}, as well as other CVs, have been reported before (see \citet{wat86} and references therein). This anomaly is associated with blendings with neighbouring lines, red-blue wing asymmetries and any deviations from the assumed flatness of the underlying continuum. Any of these reasons can shift $\gamma$ from its correct value and our spectra show that most of them apply in our case. The results from the diagnostic diagrams are given in Table~\ref{t_diag_v2051oph}. It also includes the phase offset $\phi_{0}$ and the separation of the gaussians d in \AA. Only \ion{He}{i} $\lambda$5876 has not been included in the Table because the diagnostic diagrams showed inconclusive results.

Even though the method developed by \citet{sha83} gives the primary's velocity in close approximation to its real value, \citet{mar88b} warned against the risk of biases and the fact that the value obtained might depend on the signal-to-noise ratio. Marsh proposed a different method that is less prone to bias and that takes the increased sensitivity into account. The application of this method to all the emission lines from our data set gives $K_1=89.19\pm13.70\,\rm km\,\rm s^{-1}$. If the weaker lines are not included, as in our previous $K_1$ calculation, we find $K_1=84.89\pm12.22\,\rm km\,\rm s^{-1}$. This last value is in excellent agreement with the one predicted by \citet{bap98}. The weights were calculated from the equivalent width of each line, as shown in Table~\ref{t_lin_v2051oph}.

Taking into account each line's $\gamma$, we constructed the  spectrograms of Fig.~\ref{f_spectro_v2051oph}. All spectra were folded on phases 0--1 and repeated over two cycles. In order to improve and enhance the features of the spectrograms, the spectra were binned by 4 pixels which gave optimal results. All lines show the characteristic double-peak structure, as well as the blue wing being eclipsed prior to the red one, characteristic of a prograde rotation of a disc-like emission region. The asymmetry of the emission lines is evident, in some cases though being more prominent than others. In most of the spectrograms, there is a brightness enhancement of the red wing during eclipse, probably indicating a BS origin.

Moreover, as seen from the spectrograms but even more clearly from Fig.~\ref{f_dp_v2051oph}, the double-peak separation is highly phase dependent. This is a strong signature that the disc is highly asymmetric with a non-uniform emissivity. Through period analysis, it has been found that a periodicity of $\rm P_{\rm orb}/2$ best describes the variation of the double-peak separation. This applies to all lines except \ion{Fe}{ii} $\lambda$5169, probably due to the high scatter owing to its low strength. Such a variation points to the existence of two emission sources on the AD. The indirect imaging technique of Doppler tomography, as presented in the next section, will address this issue.

%%%%%%%%%%%%%%%%%%%%%%%%%%%%%%%%%%%%%%%%%%%%%%%%%%%%%%%%%%%%%%%%%%%%%%%%%%%%%
      \begin{figure}
	\centering
	\includegraphics[width=9cm]{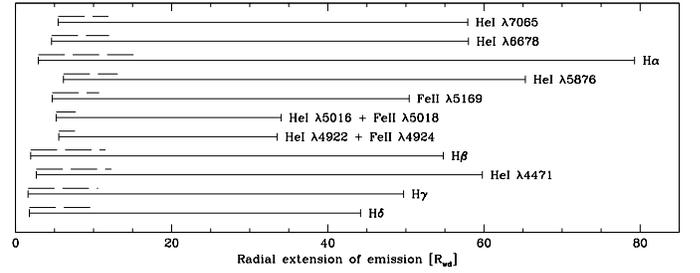}
	\caption{Radial extension of all emission lines. Dashed lines correspond to the refined outer radii when the tidal effect is taken into account.}
	\label{f_radii_v2051oph}
      \end{figure}
%%%%%%%%%%%%%%%%%%%%%%%%%%%%%%%%%%%%%%%%%%%%%%%%%%%%%%%%%%%%%%%%%%%%%%%%%%%%%     

%%%%%%%%%%%%%%%%%%%%%%%%%%%%%%%%%%%%%%%%%%%%%%%%%%%%%%%%%%%%%%%%%%%%%%%%%%%%%
      \begin{figure*}
	\centering
	\includegraphics [width=18cm]{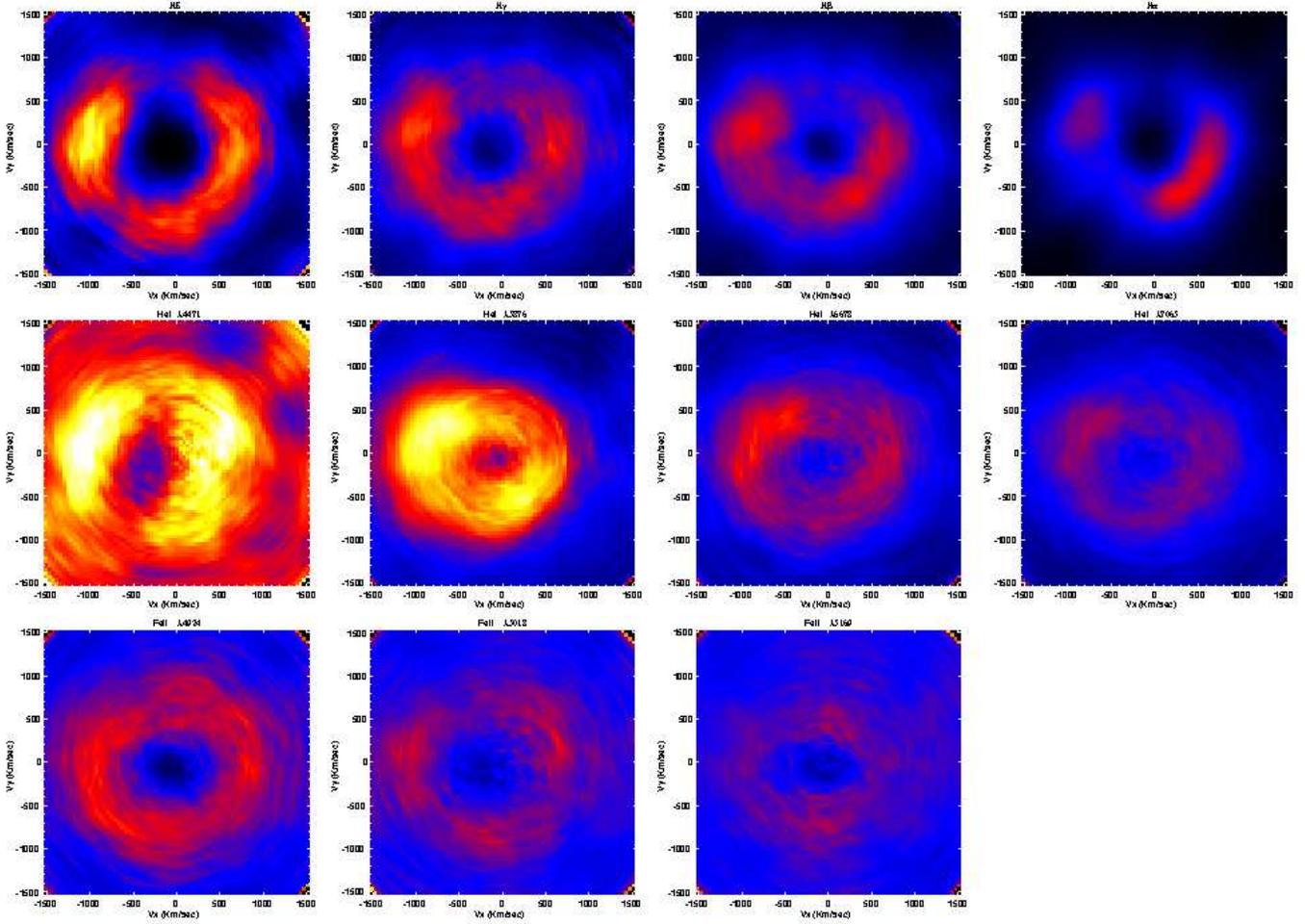}
	\caption{\object{V2051 Oph} Doppler maps. Top to bottom rows correspond to the Balmer lines, \ion{He}{i} lines and the \ion{Fe}{ii} triplet, respectively. The x and y axis represent the $V_{x}$ and $V_{y}$ velocities respectively, spanning from $-$1500 to 1500$\rm km \rm s^{-1}$. Each row has its own intensity scale. }
	\label{f_maps_v2051oph}
      \end{figure*}
%%%%%%%%%%%%%%%%%%%%%%%%%%%%%%%%%%%%%%%%%%%%%%%%%%%%%%%%%%%%%%%%%%%%%%%%%%%%%

Assuming all line emitting regions are located on the AD (however this might not be true for all lines and will be discussed in the next sections) and that the velocity field of the AD is Keplerian then the orbital radius $R_{kep}$ corresponding to a velocity $V_{kep}$ is given by the equation, $R_{kep}=GM_{wd}/V_{kep}^2$. Given the fact that the double-peak separation reflects the velocity of the gas on the outer edge of the AD and that the extension of the emission line wings reflects that of the inner disc, then the outer and inner radii of the AD are given by the equations:
\begin {equation} 
R_{in}=GM_{wd}/(V_{FWZI}/2*sini)^2)
\end {equation}
\begin{equation}
R_{out}=GM_{wd}/(V_{sep}/2*sini)^2)
\end{equation}  
where $V_{sep}$ and $V_{FWZI}$ are the velocities measured from the double-peak separation and the full width at zero intensity and {\it i} is the system's inclination. By applying the system's parameters $M_{wd}=0.78 M_{\sun}$, $R_{wd}=0.0104R_{\sun}$ and $i=83.2\degr$ \citep{bap03} we can derive the radial extension of all emission lines. The results, are demonstrated by the solid lines in Fig.~\ref{f_radii_v2051oph}. The radii are measured in terms of the white dwarf (WD) radius $R_{wd}$. From this figure it is evident that the inner radius of the disc emission for all lines is rather close to the WD, with the Balmer lines being the ones closest $R_{in} \approx 2\pm0.5\,R_{wd} $ and the rest appearing further out at $R_{in} \approx 5\pm1\,R_{wd} $. The outer radii seem to show a greater variety of behaviour, but  they are most likely overestimated since they give values greater than the volume radius $R_{L}$ of the primary's Roche lobe $R_{L}=32\,R_{wd}$. In this case tidal interaction between the secondary and the disc material will make the disc velocities sub-Keplerian and constrain the AD within the primary's Roche lobe. If the tidal effect, $\omega \rm r$, is taken into account then the refined $R_{out}$ give radial extensions constrained well within $R_{L}$. These are represented by the dashed lines in  Fig.~\ref{f_radii_v2051oph}. In this case, the outer disc revealed by H$\alpha$ is at  $R_{out}  \approx 16\,R_{wd}  \approx 0.5 R_{L}$. This is a minimum estimate of the size of the AD and not unlikely given the state of the system.

\section {Doppler Tomography}

The indirect imaging technique of Doppler tomography, first introduced by \citet{mar88}, translates orbital phase-dependent line profiles into a two dimensional velocity space map. In this way, the AD emission distribution (ring shaped structure and/or spiral patterns) and any isolated emission sources (secondary irradiation, BS \& stream trajectory) can be readily identified. For our set of data, the Doppler maps have been produced using the implementation code of \citet{spr98}, which applies the Maximum Entropy Method (MEM).

The resulting Doppler maps of the 11 selected emission lines are shown in Fig.~\ref{f_maps_v2051oph}. Both binned and unbinned spectra were used as an input but there was no significant change in the resulting doppler maps. Therefore, the unbinned spectra which gave maps of higher resolution, were adopted. The first row of Fig.~\ref{f_maps_v2051oph} shows the Balmer series, the second one the \ion{He}{i} lines and the third one the \ion{Fe}{ii} triplet. In all maps the x- and y-axis represents the $V_{x}$- and $V_{y}$-velocities, in the range of $-$1500 to 1500 $\rm km \rm  s^{-1}$. The maps of each row have the same intensity scale, while each row has a different one. This was done in order to both highlight the features seen and be able to directly compare lines of the same series.

The Balmer and \ion{He}{i} maps show the ring-shaped emission characteristic of an AD -- the result of the double-peaked emission lines, in accordance to the two maps (H$\beta$ and \ion{He}{i} $\lambda$4471 presented by \citet{kai94}. However, the higher resolution maps presented in this work have much more to reveal.} They show BS emission, albeit of different strength in each case. This emission is located at the lower part of the second quadrant ($-\rm V_{\rm x}$, $+\rm V_{\rm y}$), and was expected since the simultaneous photometric light curve, mentioned in the introduction, shows a prominent hump caused by the BS. The existence of a BS was also seen in the spectrograms of Fig.~\ref{f_spectro_v2051oph}, where there is a very bright red-shifted emission during the eclipse. This occurs when the BS emission is most obvious since most of the disc is eclipsed but the BS which precedes the secondary is fully visible. What is also evident from both  Balmer and \ion{He}{i} lines, is that the strength of the BS increases when moving to shorter wavelengths. It is most prominent in H$\delta$ and fades away when moving to lower series. This can be interpreted in terms of optical thickness. The BS appears to be optically thick in H$\alpha$, compared to the optically thin disc emission and progressively less in the rest of the Balmer series. The same is valid for the \ion{He}{i} lines. This behaviour has been observed before in other DN as shown for VW Hyi in \citet{smi06} and references therein. The map of \ion{He}{i} $\lambda$4471 is not of the same quality as the rest due to its proximity to the continuum, but has been included since it still shows the same features. 

But, the most striking feature of all, is the one seen in the $4^{th}$ quadrant ($+\rm V_{\rm x}$, $-\rm V_{\rm y}$)  of the maps, with a surprising persistence in all Balmer and He lines. Such a feature does not coincide with the location of a spiral arm, always seen in the first ($+\rm V_{\rm x}$,$+\rm V_{\rm y}$ ) and third quadrant ($-\rm V_{\rm x}$,$-\rm V_{\rm y}$ ).
%, or stream overflow. In the last case one would expect either the BS to be absent or the location where the stream re-impacts the disc to be much stronger than the BS itself. An occurrence of such a pattern has been reported in VW Hyi, albeit only tentatively. 
The existence and persistence of such a rare feature adds a great deal of interest to its origin and so will be discussed in Sect. 5.2. However, a noticeable difference in the feature's location on the different maps has to be mentioned. This is mostly seen in the Balmer series but not in the He lines. A possible trend of the feature moving downwards on the map with increasing wavelength, is visible. Its opening angle seems to differ as well, being greater in H$\alpha$ compared to H$\delta$.

The \ion{Fe}{ii} triplet shows a different behaviour than the rest of the lines. The BS is absent in the $\lambda$5169 line, and most probably in the other two lines as well. The only reason for a more enhanced emission at the BS region could be the blending with the \ion{He}{i} lines. The $4^{th}$ quadrant emission seen in the rest of the lines is absent. The difference of these maps compared to the rest is further discussed in Sect. 5.1.

\section  {Discussion}

\subsection{The iron triplet}
As briefly mentioned in Sect. 3, the three iron lines composing the iron multiplet \ion{Fe}{ii}\,42 show some distinctive differences from the other lines: (i) As seen in the average spectrum of Fig.~\ref{f_avgsp_v2051oph} they have a stronger red than blue peak, the only exception being \ion{Fe}{ii} $\lambda$4924 which seems to be the one most affected by the blend with the \ion{He}{i} line. This means that in contrast to the rest of the emission lines \ion{Fe}{ii} seems to have a greater concentration on the red part of the disc. (ii) From the same figure it is evident that the \ion{Fe}{ii} lines have a deeper central absorption as opposed again to the other lines. %This points to a region of greater optical thickness. 
(iii) The radial extension (see Fig.~\ref{f_radii_v2051oph}) also differs, with \ion{Fe}{ii} lines being emitted from an inner radius greater than the Balmer lines but similar to the \ion{He}{i} ones. The outer radius in at least two of them seems to be much more confined. (iv) The spectrograms (Fig.~\ref{f_spectro_v2051oph}) and the Doppler maps (Fig.~\ref{f_maps_v2051oph}) do not show any clear BS emission. The two blended lines do appear to have a weak red-shifted emission during eclipse but that is not seen at all for the unblended \ion{Fe}{ii} $\lambda$5169. (v) The Doppler maps differ from the ones of the other emission lines. The enhanced emission region, seen in all other lines, is absent. %and the BS is either weak (due to \ion{He}{i} blending) or absent ($\lambda$5169)
However a ring-shaped emission is clear, suggesting a  similar shape of the emitting region. 

We think that the \ion{Fe}{ii} emission can be adequately explained as being produced through the fluorescence mechanism, in which UV photons are emitted by a hot source and then absorbed by iron atoms. The absorption of \ion{Fe}{ii} in the UV triggers its emission in the optical. Absorption bands in the UV, due to \ion{Fe}{ii} have already been observed for \object{V2051 Oph} by \citet{sai06} and this iron curtain could simply be the UV counterpart of the optical emission. \citet{sai06} also suggested that this could originate from absorption by an extended gas region above the disc. The same explanation has already been given by \citet{mas05} for the case of a new short orbital period CV, \object{H$\alpha$0242-2802}. They inferred an orbital period of 107\,min and an inclination of $i\approx 82\degr$, both parameters being similar to those of \object{V2051 Oph}. However, the lack of UV spectroscopy has not yet confirmed their interpretation. It has to be noticed, that the \ion{Fe}{ii} emission lines in their spectra, covering 4200--5600\AA, also showed deeper central absorption than the Balmer lines and a stronger red than blue peak in contrast to the Balmer lines. This is in accordance with what we have observed in our spectra. In their case \ion{He}{i} emission was weak and the \ion{Fe}{ii}\,42 observed was considered to be mostly \ion{Fe}{ii} emission. 

Furthermore, \object{V2051 Oph} has not shown any \ion{Fe}{ii} emission during quiescence \citep{wil83} or in decline from a normal outburst \citep{ste01}. In consequence, its emission is likely related to the shortly terminated superoutburst, whose increased temperature may have excited \ion{Fe}{ii} in the gas region. The same has also been observed for WZ Sge, which showed no emission during quiescence but after an outburst (for more details see \citet{mas05} and references therein).  

Given all the aforementioned facts, the most plausible explanation is that \ion{Fe}{ii} does not come from the same line-forming region as the Balmer and \ion{He}{i} lines, therefore does not form on the AD itself but from an extended gas region above it.

\subsection{The $4^{th}$ quadrant emission}

In the Doppler maps of Balmer and \ion{He}{i} lines we have seen an arc-shaped region of enhanced emission located on the $4^{th}$ quadrant. It should be noted that the existence of two locations of prominent emission in the maps was already evident from the modulation of the double-peaks separation (Fig.~ \ref{f_dp_v2051oph}). Since one of them is the BS and the double-peak separation is modulated with half the orbital period, that means that the second region of emission should be opposite the BS, as seen in the maps.

Such emission has also been reported by \citet{smi06} for another SU UMa star, \object{VW Hyi}, just prior to the onset of an outburst. However in this case the emission was only seen in the Doppler map of H$\alpha$ and remained absent in the rest of the Balmer and \ion{He}{i} lines (they shared a similar spectral coverage to ours). The appearance of this feature in H$\alpha$ maps alone, and their limited phase coverage, left them unconvinced of its existence. 

We recall that we have captured the system shortly (2\,d) after the end of a superoutburst and that it showed photometric superhumps at the time of our observations  \citep{pat03}. It is then quite natural to assume that the enhanced emission region seen in our maps is nothing else but the equivalent of the superhump light source (SLS). Concerning its shape, for the majority of systems, the SLS is an extended region of the disc involving $\approx 1/4$ of its area (\citet{war95}, p.199), which agrees with our maps. \citet{war87} saw a narrow hump in their light curves during outburst which they associated with a possible superhump. They gave phase 0.54 for its emission peak. Such an emission should be seen at the lower half of the $4^{th}$ quadrant in our maps. In H$\beta$, H$\alpha$ and \ion{He}{i} $\lambda$5876 this could be true, while it does not appear to hold for H$\delta$, H$\gamma$, or \ion{He}{i} $\lambda$4471, for instance. However, in any cases, caution should be drawn to such an interpretation, since \citet{war87} did not fully trust the reality of their signal due to rapid variations of brightness at that time.

Additionally, as pointed in Sec.~1 the system at that time had a different emission distribution in its disc. The V/R variations of their spectra were in almost anti-phase with ours and their average spectrum showed a stronger blue-to-red peak for H$\alpha$ in contrast to ours.

Still, the physical origin of the SLS for CVs is unclear. As briefly mentioned before, this emission's origin most probably should not be attributed to stream-disc overflow. SPH simulations by \citet{kun01} offer additional support to this opinion. In their study they used, among others, the parameters of the SU UMa star \object{OY Car}, which is considered the ``twin'' of \object{V2051 Oph}. Their 3D simulations showed that the re-entry region of the overflowing stream onto the disc is located close to zero velocity, at orbital phase 0.5. Such a region of enhanced energy is best pictured by Doppler tomography and in our maps should be expected on the border of the $3^{rd}$ and $4^{th}$ quadrants. However, the enhanced emission region in all our maps (except perhaps for \ion{He}{i} $\lambda$5876) does not agree with the one predicted by the numerical simulations (see e.g. references in \citet{kun01}). We caution however for the fact that our data set gives us only a resolution in the maps of about 33$^\circ$. It is thus difficult to determine the boundary between quadrants precisely and reach a definitive conclusion.

However, if this feature is indeed associated with the SLS, it is rather puzzling that it was seen at the onset of the outburst of \object{VW Hyi}, when the disc was not large enough for superhumps to develop. 

Another possibility is that the arc-shaped emission is due to tidal interactions of the secondary and the disc, like the two spiral arms observed in DN in outburst. But this has to be ruled out since simulations predict and observations verify that the two spirals are seen in the $1^{st}$ and $3^{rd}$ quadrants, which is not the case.

\section {Summary}
The results of this study can be summarised as follows:
\begin{enumerate}
\item The average spectrum, covering 4000--7500\AA, shows Balmer, \ion{He}{i} and \ion{Fe}{ii} lines in emission. \ion{He}{ii} also appears in weak emission. All lines except the \ion{Fe}{ii} ones show a stronger blue peak. The lines on the blue part of the spectrum show a red-blue wing asymmetry.
\item The Balmer decrement from H$\alpha$ to H$\delta$, 1:0.67:0.56:0.43, shows a mix of optically thin and thick conditions.
\item The primary's velocity was found to be 84.9\,$\rm km \rm s^{-1}$, in agreement with previous estimated values.
\item The trailed spectra clearly demonstrate the blue wing being eclipsed prior to the red one, characteristic of the prograde rotation, as well as the double-peaked structure, the line asymmetries and the red wing brightness enhancement during eclipse (except the \ion{Fe}{ii} lines) characteristic of BS emission.
\item The double-peak separation of all emission lines is modulated with half the orbital period, indicating a highly asymmetric disc and the existence of two emitting sources.  
\item The disc emission for all lines originates from close to the WD to $\approx \rm R_{\rm L}/2$. Only the \ion{Fe}{ii} lines' radial extension is more constrained.
\item The Doppler maps of the Balmer and \ion{He}{i} lines confirm the expected ring-shaped emission of the AD and the BS emission, which increases with decreasing wavelength. This indicates a progressively less optically thick BS (compared to the optically thin disc emission) when moving to shorter wavelengths. A striking arc-shaped enhanced emission region is also seen in the $4^{th}$ quadrant. This probably represents the SLS. Its physical origin though is quite uncertain and cannot be attributed to stream-disc overflow or spiral arms as observed in DN in outburst. 
\item In contrast to the rest of the lines the Doppler maps of the \ion{Fe}{ii} triplet does not show any BS emission (except some hints in the two lines blended with \ion{He}{i}) or the $4^{th}$ quadrant enhanced emission. We conclude that the \ion{Fe}{ii} lines originate in a gas region above the disc and are produced through the fluorescence mechanism. Furthermore the \ion{Fe}{ii} emission must be associated to the recently terminated superoutburst, during which the increased temperature excited the \ion{Fe}{ii} in the gas region. Previous studies during quiescence or decline from normal outburst did not show any \ion{Fe}{ii} emission.  
\end{enumerate}
Additional follow up through high-resolution time-resolved spectra and application of Doppler tomography, of other SU UMa stars and especially this one is called upon. Catching the system at other brightness states and using a similar observational setup will also allow us to see if this feature persists and how it behaves. It is evident that a lot of information can be gained by the simultaneous study of several emission lines. SU UMa stars are short period CVs, and the use of even larger telescopes will help us achieve an even higher phase resolution. What would also be beneficial, is for the system to be monitored for several days. In this way, the precession period $\rm P_{\rm prec}$ of the AD could be detected. For \object{V2051 Oph} it will be around 2.3 days, as estimated from the superhump period.

\begin{acknowledgements}
  We kindly thank C. Tappert for providing the MIDAS adapted version of the Doppler tomography and diagnostic diagrams codes. The authors wish to thank the anonymous referee for pertinent remarks on the manuscript. C.P. wishes to thank A. Bonanos for useful comments on the manuscript and gratefully acknowledges a doctoral research  grant by the Belgian Federal Science Policy Office (Belspo). 
\end{acknowledgements}
\bibliographystyle{aa}
\bibliography{references}

\end {document}